\documentclass[11pt]{article}

\begin{document}

\title{\vskip-2.5cm {\sf When will fusion energy truly become a reality?} }
\author{{\bf Renato Spigler$^{1,2,*}$}   
                 \\  \\
   $^1$Department of Mathematics and Physics, Roma Tre University
                 \\
          1, Largo S. Leonardo Murialdo, 00146 Roma, Italy
                 \\
                $^*${\em Program IGNITOR}
                 \\  \\
           {\tt renato.spigler@uniroma3.it} }


\maketitle

\begin{abstract}
This abstract provides up-to-date insights into fusion (thermonuclear)
research, detailing ongoing projects and planned devices. The document also
explores alternative sources of energy, offering a comprehensive overview of
the current landscape. Additionally, notable comments and observations are
provided to illuminate key aspects of the discussed topics. Stay informed as
we delve into the latest advancements and initiatives in the dynamic field of
energy research.
\end{abstract}
       
\section{Introduction}

The demand for energy, in increasingly greater quantities, appears to be
growing more clearly (and unfortunately dramatically) today, both from
industrialized countries and so-called emerging ones. In fact, the former need
energy not only to improve their quality of life but also to meet the needs of
increasingly sophisticated and necessary industries and services (well-being,
medicine, communications, transportation). However, energy is also needed by
emerging countries (third and fourth world), often highly populous, which are
following the path previously taken by today's more developed countries. These
countries, for various reasons, tend to exploit polluting forms of energy
derived from fossil sources.

Apparently, today there is, and indeed it is growing in more advanced
countries, a greater sensitivity towards the environment, and thus a tendency
to resort to clean and safe sources of energy. However, it is difficult to
outright condemn poorer countries that have fossil resources but cannot afford
the costs associated with the development and adoption of sustainable sources.
It is not to say that these countries are devoid of concern for the
environment, but the need to survive, even in conditions endangering public
health, outweighs the prudence required in the use of energy sources that the
West strongly criticizes today and would like to abolish altogether.

While, to the chagrin of staunch environmentalists, so-called alternative
sources of clean energy offered by nature (hydroelectric, wind, solar, marine,
biomass, geothermal) seem to make a very modest contribution, insufficient for
the needs of both developed and developing societies, nuclear energy remains on
the table, both from fission reactors (existing) and the one pursued for a long
time (although this delay has been caused by unjustified deviations), coming
from fusion.

The environmental cost, as well as upstream and downstream pollution and
storage difficulties related to various forms of alternative energy, should
also be taken into account. Furthermore, these are linked to climatic
conditions, which vary in different regions of the country, whereas a nuclear
power plant of any kind would produce a negligible amount of waste, although
they do present potential hazards to be assessed.

\section{Fission Energy}

Fission energy is still viewed today as dangerous, both for the radioactivity
that would accompany it in the event of natural disasters or human errors and
for the radioactive waste that would remain hazardous for very long periods.
However, there are now several "generations" of fission reactors considered
progressively safer, and perhaps the disasters recorded so far (Three Mile
Island, Chernobyl, Fukushima) have been overestimated. It has also been
proposed to place fission reactors underground, at depths of 200-300 meters.

The average person is not accustomed to "living by statistics" and may not be
aware, for example, that the COVID-19 vaccine, opposed by some, carries a
significantly lower risk than being struck by lightning or dying from an
anaphylactic shock caused by an insect bite.

The most educated individuals and the political class of a modern country,
which should ideally be its emanation, should take on the responsibility of
{\em properly informing} the population. Not all decisions can be left to the
entire population because specific expertise in every field is crucial.

It is a fact that there is widespread misinformation today, not only in Italy
but also in other countries, the spread of incomplete, incorrect, or
deliberately biased information, sometimes by ignorant individuals, often by
incompetents, and other times for very specific vested interests. The results
are always damaging.

For many years, we have known that splitting atoms (easier if they are heavy
like uranium and plutonium) releases energy. Unfortunately, the uncontrolled
version of this process has given us the atomic bomb, while the controlled
version has provided peaceful energy through appropriate reactors. This is
fission.

Today, it seems that a fourth generation of fission reactors, considered safe,
is on the horizon. We will see. Meanwhile, even the less safe ones are a source
of profit for countries like France and [until now] Russia, which build and
sell them around the world. Does it make sense for Italy not to want them on
its soil but to purchase fission energy from France, Switzerland, and Slovenia?
Would a potential catastrophe beyond the Alps not spread radioactivity here as
well? Or would it stop at the Alps due to a referendum?

\section{Fusion Energy}

Fusion energy has been discussed for a long time, and at the moment, there is
renewed emphasis on it. What does it consist of? We know that even by "fusing"
atoms (easier if they are light like hydrogen and its isotopes, deuterium and
tritium, or helium), energy is released. Again, we have seen a military
application of this in the so-called hydrogen bomb or "H-bomb." But once again,
as evidence that the use of the results and products of science and technology
depends on responsible human choices, nuclear fusion reactions also have a
possible peaceful application. This is the path that we would like to follow by
achieving controlled thermonuclear fusion in a "fusion reactor." This, given
the absence of inconveniences from reaction products, for example, by using
appropriate mixtures of deuterium and helium-3, as well as - in principle - the
wide availability of material to use as fuel, would represent the solution to
all our problems: a virtually inexhaustible source of sustainable energy. Not
insignificant. It should be noted that while deuterium can be easily obtained
from water (through distillation), helium-3 is less readily available (it can
be obtained from deuterium-deuterium reactions or brought from the Moon, where
it is abundant, although this is not currently feasible).

In relation to fusion, it is rightly said to be energy similar to that produced
in the Sun. All stars are "kept alive" by thermonuclear reactions, the same
ones that occur in the aforementioned fusion. Stars will "die," cease to exist
as such, when all their fuel is exhausted, i.e., when there are no more light
elements to use for fusion. However, the time required for a star to burn all
its fuel is typically measured in billions of years, and it is estimated that
our Sun still has 4 or 5 billion years to live.

Achieving fusion on Earth, however, presents non-trivial difficulties. The
physical state in which matter appears in stars is that of plasma, i.e.,
ionized gas: its atoms are divided into ions (positively charged) and electrons
(negatively charged). This gas, as one can imagine, tends to expand, to escape,
which would prevent the collision of particles (more precisely, ions) that, by
doing so, fuse to create heavier ions, while simultaneously releasing fusion
energy. In stars, whose mass is considerable, it is gravity that confines the
gas, but on Earth, this is not the case. It's not enough: the necessarily
modest amount of ionized gas produced in the laboratory cannot be confined by
the gravitational attraction of the Earth.

In principle, there are fundamentally two possible solutions to achieve fusion
reactions in a laboratory and, hopefully, in a real reactor. One is based on
the fact that ionized gas particles are electrically charged and therefore
sensitive to the actions of electric, magnetic, and electromagnetic fields.
This leads to the conception of, for example (but not only), "toroidal"
devices, shaped like a doughnut or a "torus," as mathematicians say, in which
charged particles are confined because they are forced to orbit around the
lines of the magnetic field created within the "torus." However, the confined
plasma must have a sufficiently high density and temperature and remain
confined for a sufficiently long time for there to be a sufficiently high
probability of fusion reactions taking place. Achieving these conditions in the
laboratory is not straightforward.
For a long time, it was believed that building larger machines would produce
the desired result, but unfortunately, this does not seem to be true, while the
cost of building a toroidal machine increases drastically with its size.

A machine of a certain size designed to conduct fusion experiments is the JET
(Joint European Torus), located at the Culham Centre for Fusion Energy in
Culham, Britain. JET has been operational for many years, from 1983 to the
present, and will cease operation by the end of 2023. It has produced several
experimental results by numerous researchers from around the world, but it
should be clear that JET was never conceived as a prototype reactor, i.e., a de
ice capable of achieving ignition, initiating a thermonuclear reaction, and
sustaining the reaction, producing more energy than needed to start the process.

{\em It is a fact that the currently partially constructed ITER machine (in
Cadarache, France), widely publicized in the press, does not, and cannot, as
pe the current design, have the objective of achieving ignition}, let alone
building a thermonuclear reactor within a couple of decades.

A second way to achieve fusion, distinct from magnetic confinement, is through
inertial confinement. In brief, this involves inducing fusion by extreme
compression of a mixture of deuterium and tritium using various devices and
processes, such as a high-power laser beam or high-energy particle beams.

\section{What happens in the Sun?}

Nuclear fusion reactions responsible for the Sun's energy production occur in
its core, where temperature and density are higher. In the Sun's core, hydrogen
is converted into helium (via deuterium). As of today's knowledge, the Sun's
core is predominantly composed of hydrogen. The temperature is around 16
million degrees Celsius, the pressure is extremely high, around 500 billion
atmospheres, and the material's density is approximately 150,000 kg/m³. These
conditions are exceptional on a human scale, i.e., on Earth, but they are
normal in a star.

At these temperatures, hydrogen atoms in the Sun's core cannot remain intact
and split into protons and electrons. The thermal energy is so high that when
protons randomly encounter each other, they overcome the electrical repulsion
between charges of the same sign and merge to form a helium nucleus.
Approximately 594 million tons of hydrogen fuse every second, releasing energy
equivalent to 386 trillion trillion megajoules. This energy is equivalent to
the mass of 4 million tons of hydrogen, while the remaining 590 million tons
are converted into helium. Consequently, our Sun loses 4 million tons every
second, but its overall mass is so large that even after about 5 billion years
of active life, it has only slightly reduced its mass.

In summary, fusion reactions involve the nuclei of light elements like hydrogen
fusing into heavier element nuclei like helium at high temperatures and
pressures, as mentioned above. It should be noted that hydrogen exists in
various forms (isotopes) in the Sun's core, including regular hydrogen (H),
deuterium (D), and tritium (T). In the D-T thermonuclear reaction, for example,
between a deuterium nucleus and a tritium nucleus, a helium nucleus (alpha
particle) and a neutron are generated.
The total mass of the reaction products is less than the sum of the masses of
the interacting particles, which is why energy is released according to the
mass-energy equivalence principle.

\section{But can fusion be achieved on Earth?}

Decades of experiments have alternately generated the illusion of imminent
success and the discouragement of not being able to achieve fusion in a
laboratory on Earth. However, this is not the opinion of the international
scientific community.

Agreeing with Niels Bohr, it is not easy to make predictions, especially about
the future, but there are serious indications of what may happen regarding
fusion. While the common press occasionally reports, but recently quite
frequently, that China has reached exceptional temperatures for extended
periods (but with plasmas too sparse, and the journalist does not specify), or
that a machine (yet to be built) will bring "the Sun into our homes" within a
few years, or that the private English company Tokamak Energy has reached 100
million degrees in an experiment conducted on a spherical tokamak (without
specifying plasma density and confinement times), the outcome of an experiment
conducted at the National Ignition Facility (NIF), located at the Lawrence
Livermore National Laboratory in California, announced in August 2021, seems to
have received little publicity. In fact, here, using the inertial confinement
method, there has been substantial evidence that the aforementioned ignition
has indeed been achieved. The result is exceptional because it demonstrates, at
least, that {\em nuclear fusion is achievable in a laboratory on Earth}.

What about fusion in magnetic confinement machines? After the European JET
initiative, the construction of a larger machine, ITER (International
Thermonuclear Experimental Reactor), was scheduled in Cadarache, France, in
2007. According to what is read on Wikipedia, {\em ITER, also intended in the
original Latin sense of "path" or "journey," is an international project aimed
at building an experimental nuclear fusion reactor capable of producing a
fusion plasma with more power than is required to heat the plasma itself.}

From what has been subsequently declared by the designers themselves (the
machine is still in the construction phase, but the design itself does not
seem to have been completed), it does not appear that ignition is among the
achievable goals of ITER.

If this is the case, in the face of a construction
phase lasting at least three decades (the scheduled year for the start of the
first plasma experiments has currently been pushed to December 2025), and an
expenditure that has risen from 5 to 10 billion to 60-65 billion Euros or more,
we may end up with a machine that could provide less than JET! This is all the
more serious when you consider that ITER should be followed by a true prototype
of a thermonuclear reactor, named DEMO (DEMOnstration Power Plant).

{\em It is worth noting that if ITER were to fail, the enormous expense and
time invested would likely lead almost all countries in the world to halt
significant funding for fusion research for who knows how long.} The economic
damage would be enormous.

Over the years, while the USA have substantially reduced its funding for the
ITER program, many countries, through political figures and members of the
scientific community, have openly supported it.

In light of the previous criticisms, the strong support for ITER seems somewhat
inexplicable, unless it is due to a limited or outdated scientific vision or
economic interests, such as favoring the sale of fission reactors (delaying
fusion) or benefiting industries present in various countries, but all of this
has nothing to do with the realization of a true fusion reactor.
The fact is that the validity of a different paradigm from that followed by
ITER has been consolidated by several parties, namely that of compact machines,
relatively small in size and with high field strength, characterized by
particularly high magnetic confinement fields. Ignitor is an example of such a
machine, based on an Italo-Russian agreement, also involving the USA and signed
some time ago. It certainly does not aim to be a reactor, but one of its
objectives is the concrete possibility of approaching ignition. To confirm
this, the machine should be built, and its core is estimated to cost no more
than about 78 million euros (to be compared with the total estimated cost of 65
billion euros or more for ITER, which does not even have the goal of ignition).
Why does Ignitor cost so little? First of all, the sites designated to host the
machine are already available (in Caorso, while initially Troisk in Russia was
considered), the design cost has already been almost entirely paid, and many
people working on it do so without pay, as they are researchers and university
professors who already receive a base salary. The 78 million euros, which were
allocated to Ignitor (and after these years of waiting, they can be considered
not more than 100 million), are used for the construction of the machine's
core, i.e., the machine itself, without considering the infrastructure. Its
various parts must be built by Italian industries, with a small part of the
cost allocated to updates and design checks by qualified professionals
(engineers) in Italy.

Ignitor {\em is a scientific experiment} based on a machine whose construction
is possible, albeit with numerous bureaucratic obstacles, and it is part of an
agreement signed by Italy and Russia in 2011 (with the approval of the USA),
during the time of Berlusconi and Putin. The agreement envisioned the
construction of various parts of the machine by Italian industries, after which
these parts would be shipped to Russia and assembled in Troisk, where there is
a suitable site to host and power the machine.

The dramatic events related to Russia's invasion of Ukraine have likely
strained relations with Russia, even in the scientific field, and this will
likely last for an indefinite period. However, the urgent need to meet energy
demands should push our country to build Ignitor in Italy. Moreover, the idea
and project are the work of Bruno Coppi, an Italian professor at MIT in Boston,
and Italian industries that have always been planned to build various parts of
Ignitor. Therefore, with a more than suitable site like Caorso, there is no
reason to delay. We don't need Russia to move forward.

But why was such a modest amount as the one requested by Ignitor - the 78
million euros allocated during Minister Gelmini's time as part of the "flagship
projects" - mostly reallocated elsewhere by Minister Fedeli, without even
consulting the project's leader, Bruno Coppi? And why has ENI been
participating in the Commonwealth Fusion Systems (CFS) consortium since 2018,
an MIT spin-off, with already 50 million euros paid, to build SPARC, a machine
that is essentially an attempt to copy Ignitor, without even consulting the
originator? All of this remains a mystery.

In addition to the {\em scientific} value of the experiment and its potential
{\em economic} benefits, with the ongoing energy crisis and over 40\% of
Italy's energy needs dependent on Russian gas, the diplomatic value of the
aforementioned Italo-Russian agreement (concluded with the approval of the USA)
should have been taken into greater account. The agreement, which had already
been signed but never fully honored, could also be interpreted as a cooperative
action in favor of nuclear peace, as this had already begun with Coppi and his
Russian counterpart, E. Velikhov. Unfortunately, this latter goal now seems to
be obsolete.

\section{The idea of ``hybrid reactors''}

There is a possibility that would allow the use of fission reactions with
reactors having more advanced characteristics than those currently built.
These are the so-called ``hybrid reactors,'' machines in which the source of
neutrons needed to produce fission reactions is a fusion machine, without the
need for the latter to reach ignition. The most desirable fissile material
would be thorium and plutonium. This innovative solution would combine the
reliability of traditional fission reactors with safety.

The remarkable fact is that, unlike fusion, {\em the technology required to
build a hybrid reactor already exists today}. Moreover, Italy would be in a
pole position in this field since the aforementioned technology follows that
already developed to build experimental machines such as Alcator A (in
operation from 1973 to 1979), Alcator C (from 1978 to 1987), and Alcator C-Mod
(from 1991 to 2016), named after the Italian "Alto Campo Toro," which operated
at the Plasma Science and Fusion Center at MIT in Boston, as well as FT
(Frascati Tokamak) and FTU (Frascati Tokamak Upgrade), which started operating
in 1977 and 1989, respectively, at ENEA in Frascati, and concerning the
aforementioned Ignitor project. It is no coincidence that Russia has already
decided to develop an experiment along these lines.

\section{But where are we really at?}

Unfortunately, for reasons sometimes known and sometimes not, it is a fact that
we are surrounded by considerable misinformation.

What is comforting is the result of NIF, which at least shows that nuclear
fusion is feasible in a laboratory on Earth using inertial confinement. Less
promising are the results obtained or expected so far with magnetic
confinement. In summary:

$\bullet$ ITER: Apart from the considerable cost and estimated completion time,
it does not even plan to achieve ignition, which does not provide much hope
that the subsequent DEMO can truly be a reactor. The Russian invasion of
Ukraine, with the consequent sanctions imposed by the EU and the USA, will
probably cause further damage to ITER.

$\bullet$ DTT (Divertor Tokamak Test): It is supposed to be realized at ENEA in
Frascati, with ENI and the CREATE Consortium, but there is still not enough
justification for proceeding with it. Regarding DTT, it has been explicitly
stated that, "at an estimated cost of approximately 600 million euros, ENEA and
ENI will build it in Frascati over the next seven years, with substantial
national funding (10\% provided by the EU), a project aimed at creating a
"diverter," a device designed to expel the energy - mostly heat - and the
products of nuclear fusion generated inside the tokamak. It would be a very
flexible machine in operational scenarios and represents a significant
advancement in terms of performance compared to machines conceived over 40
years ago. DTT's purpose is to qualify the divertor prototypes for DEMO (the
demonstrative fusion machine that would follow ITER), but its operation will
also allow the growth of new generations of scientists, capable of working on
ITER and DEMO." In reality, the cost has already increased to 650 million
euros, but it is a common opinion that after the project is completed, it will
increase to at least double that amount.

Unfortunately, these are just
hypotheses. If ITER is not completed according to the plan, the idea of DEMO
and, consequently, the DTT, which would serve it, would be of little use. Only
the fact that "new generations of scientists" can grow, as happened with JET,
working on fusion scientific experiments remains. But at what cost to
taxpayers? In reality, continuing down this path will leave Italy without
significant experiments for who knows how many years, such as FT and FTU at
ENEA in Frascati, which have operated for decades.

$\bullet$ SPARC, the machine of the aforementioned Commonwealth: it tries to
reproduce (well?) the machine already designed for the Ignitor Program, but for
the moment, only one of the numerous coils necessary to form a superconducting
coil has been built. It is not even clear if there is enough of the required
materials worldwide, as they are very rare elements. While Ignitor would use
superconducting MgB2 ceramics, it would be much more problematic to use ReBCO
superconductors (barium and rare earth copper oxide).
What is more realistic is
to hope for the development of new superconducting materials that are more
accessible, for which there have been recent developments at Princeton.

$\bullet$ JET: The machine, still in operation at Culham, has produced widely
publicized results. Unfortunately, these are not as remarkable as claimed: far
more than 30-35 MW of power has been injected to achieve a plasma heated to
only 10 MW. Furthermore, it should be clarified that only one-third of this
power can be considered directly ``usable.''
Anyone understands that to gain an
advantage, the ratio between the useful power obtained and the power supplied
must be greater than 1.

The BBC has raised some doubts about the touted
success, and indeed, an informed source reportedly said that the ratio of
energy input to energy output in the JET facility has unfortunately remained
unchanged compared to the past.
Other experts have pointed out that the purpose
of the JET experiment was actually to reproduce the result (i.e., gain)
obtained 20 years ago but with higher power injection, as well as to do it with
a new inner wall made of beryllium and tungsten. This was done in preparation
for future experiments on ITER. From a technical point of view, this latter
point was interesting because it is known that carbon (graphite), used in the
past on the JET's inner wall, retains Tritium, and there are various
differences in the physics that occur in the two different conditions mentioned.

$\bullet$ Ignitor: The reason why work on its construction is effectively
hindered at various decision-making levels is not disclosed. The machine has
also been imitated (at a much higher cost) by SPARC, not to mention the
feasible development of hybrid reactors, which are based on the same technology
and existing knowledge of physics.

Regarding Ignitor, L.J. Reinders writes in his recent book ``{\em The Fairy Tale
of Nuclear Fusion}'' (Springer Nature, Switzerland AG, 2021, par. 6.5, page
165): \cite[par.~6.5, page~165]{Reinders}
``Finally, we mention the joint Russian-Italian project on the IGNITOR reactor,
which has evolved out of Bruno Coppi's activities at MIT (see Chap. 5).
It is part of the line of research that started with the Alcator machine at MIT
in the 1970s, and continued with Alcator C/C-Mod at MIT and the FT/FTU
experiments at Frascati. {\em It is so far the first and only experiment
proposed and designed to obtain physical conditions in magnetically confined
D-T plasmas that sustain the plasma under controllable conditions without the
addition of extra heat, i.e., to achieve ignition.} It is a compact D-shaped
fusion reactor with a total plasma volume of just 10 $m^3$. So far, only model
calculations have been carried out (Bombarda et al. 2004), and construction of
the reactor itself at the TRINITI site in Troisk (Russia) is long overdue.''

The observations in this article are not meant to be entirely negative,
hypercritical, or overly pessimistic, but the fact remains that only serious
{\em scientific knowledge} and {\em accurate information} are the basis for
solving and publicizing such important and no longer avoidable problems as
those of energy. Disinformation can only harm by deluding taxpayers and
directing public and private funding toward less appropriate and promising
paths. This is also because alternative solutions for obtaining gas energy do
not seem to provide large quantities or do so in the short term. On the one
hand, it is estimated that increasing gas extraction in Italy, especially
around the coasts, besides presenting well-known environmental risks, would
only produce 10\% more. On the other hand, as pointed out by Romano Prodi, the
practical use of gas extracted elsewhere, far from Italy, requires three
phases: (1) it must be transformed into liquid form for transport; (2)
transportation must be done through special gas pipelines or methane carriers;
(3) the liquid gas must then be transformed back into gas (regasification). At
the moment, there are not enough facilities or ships equipped for all of this
in Italy, and it certainly cannot be organized overnight. From the point of
view of users, however, it should not be forgotten that the cost of energy on
bills is not determined only by the cost of energy itself but also by
significant tax contributions, partly justified for social expenses.

\section{Conclusions}

It seems advisable not to risk losing, as a country-system, the opportunity to
play a central role on the international stage once again, as happened in the
sad cases of Chemistry during Natta's time and Electronics during Olivetti's
time. There are all the prerequisites for proceeding toward successful
objectives. From an operational point of view, a coordinated intervention that
includes the participation of five of the current ministries would be
advisable: the Ministry of Ecological Transition, the Ministry of Economic
Development, the Ministry of Economy and Finance, the Ministry of University
and Research, and the Ministry of Foreign Affairs and International
Cooperation, probably by creating an ad hoc agency.

\section{Acknowledgments}

The author thanks Bruno Coppi (MIT), Alessandro Cardinali, Cristina Mazzotta,
and Gianluca Pucella (ENEA-Frascati), and Valeria Ricci (University of Palermo)
for their expert observations made during the preparation of this article.

\end{document}